\documentstyle[multicol,aps,epsf]{revtex}
\begin{document}
\draft \title{Phase Ordering in Nematic Liquid Crystals}
\author{Colin Denniston$^{1}$, Enzo Orlandini$^{2}$, and J.M. Yeomans$^{3}$}
\address{$^{1}$ Dept. of Physics and Astronomy, The Johns Hopkins University,
Baltimore, MD 21218.}
\address{$^{2}$ INFM-Dipartimento di Fisica, Universit\`a di
  Padova, 1-35131 Padova, Italy}
\address{$^{3}$ Dept. of Physics, Theoretical Physics,
University of Oxford, 1 Keble Road, Oxford OX1 3NP}
\date{\today} \maketitle

\begin{abstract}
We study the kinetics of the nematic-isotropic transition in a
two-dimensional liquid crystal by using a  lattice Boltzmann scheme
that couples the tensor order parameter and the flow consistently.
Unlike in previous studies, we find the time dependences of the
correlation function, energy density, and 
the number of topological defects obey dynamic scaling
laws with growth exponents
that, within the numerical uncertainties, agree with
the value $1/2$ expected from simple dimensional analysis. 
We find that these values are not altered
by the hydrodynamic flow.  In addition, by examining shallow quenches,
we find that the presence of orientational disorder can inhibit
amplitude ordering.
\end{abstract}

\section{Introduction}

The phase ordering kinetics of liquid crystal systems undergoing 
nematic-isotropic transitions has attracted considerable experimental 
\onlinecite{CTY91,CDTY91,PGY94} and theoretical
\onlinecite{BB94,YP93,ZG95,F98} interest. 
One of the reasons is that nematic liquid crystals provide an 
experimentally accessible system with continuous symmetry that, unlike
systems with discrete symmetry, can display stable topological
defects.  During phase ordering these defects interact and annihilate 
and it is believed that many of the universal properties of the late 
stage kinetic growth can be explained in terms of the
defect dynamics. This property is also shared by the well studied 
$O(N)$ model. However here the continuous symmetry belongs to a 
different homotopy class (the $O(N)$ model lacks the inversion
symmetry present in the director field of nematic liquid crystals). 
It is not yet completely clear if such 
a difference in the symmetry can modify the ordering dynamics of the
nematic liquid crystal.

The aim of this paper is to study the kinetics of phase separation for 
two-dimensional liquid crystals under a quench from an isotropic
to a nematic phase.  Considerable controversy exists as to whether the
ordering violates dynamical scaling \cite{ZG95}.  The simplest scaling
analysis based on the assumption of diffusive dynamics for the order
parameter suggests that any length scale in the system should grow
with a power law in time $t^{1/2}$ \cite{BB94}.  However, an
examination of the configuration of the order parameter in experiment
(or simulation) clearly shows that the late stage ordering proceeds by
defects moving to annihilate.  Simple defect arguments which rely on an
assumption of a finite constant friction coefficient for the movement
of defects also give $t^{1/2}$ \cite{BB94}.  A problem arose due to
early calculations \cite{IO73,dG76,LL86} which showed that the
friction coefficient diverges logarithmically with system
size.  This brought into question whether the $t^{1/2}$ behavior from
the simple scaling analysis would be found.  Simulations of the $XY$
model \cite{BB94} and later of tensor models of liquid crystals
\cite{ZG95,F98} reinforced this view when they failed to measure the
$t^{1/2}$ behavior and, in fact found exponents which appeared to be
decreasing away from $1/2$ at late times \cite{ZG95}.  However,
experiments by Pargellis and coworkers \cite{PGY94} found behavior
consistent with the $t^{1/2}$ power law.  In addition, their
simulation of the $XY$ model with very large amplitude noise
(considerably larger than that found experimentally) 
agreed with the $t^{1/2}$ power law.  A later more rigorous
calculation of the friction coefficient for defect dynamics shows that
it does not diverge with system size, but only with vanishing defect
velocity\cite{CD96}.  This suggests that it should be possible to
measure the $t^{1/2}$ behavior in a simulation.

Another important issue which remains essentially unexplored
is the extent to which  
the presence of hydrodynamic modes coupled to the nematic order 
parameter can affect the late stage kinetic growth.
It is well known that hydrodynamic interactions can
crucially influence phase transition kinetics, for example in simple 
binary fluid mixtures and in gas-liquid systems 
\onlinecite{OOSYB95,Cates99}.  In liquid crystals
the coupling between the local molecular orientation and 
the velocity field influences the 
dynamics of the liquid crystal in a complicated manner. 
For example, shear flow will cause the molecules 
to reorient and conversely a
reorientation of the liquid crystal may induce a velocity field (backflow).
To have a complete picture of the 
kinetic growth of a phase separating nematic liquid crystal 
it is crucial to include the effect of these couplings.

 Lattice Boltzmann approaches have proved very
successful algorithms for investigating phase ordering in binary
fluids. Here we use an extension of the method which models liquid
crystal hydrodynamics\onlinecite{DO00,us00}.  
The equations of motion describing liquid crystal 
hydrodynamics are complex.  There are several derivations
broadly in agreement, but differing in the detailed form 
of some terms.  Here we follow the approach of Beris and 
Edwards who write the equations of 
motion in terms of a tensor order parameter ${\bf Q}$ 
which describes the orientational distribution function of 
the molecules\onlinecite{BE94}. This formalism is appropriate here
because the motion of defects is
explicitly included. Moreover both the isotropic and nematic phases
can be modelled using the same formalism which is necessary if the
dynamics of the transition between them is to be followed.

The paper is organized as follows.
In Section II we summarize the hydrodynamic equations of 
motion for liquid crystals. 
The lattice Boltzmann scheme used to model these equations is
described in Section III.
In Section IV we introduce the correlation function, the correlation 
length, the energy density, and the defect separation as the relevant 
quantities that characterize the late time behavior
of the phase separating system. A review and discussion of previous 
results is also given. The dynamic scaling behavior of the
model is investigated for systems of linear size $L=256$,
and the space of the relevant parameters of the model is explored to
choose the best values for more computationally intensive simulations.
In Section V a  careful estimate of the dynamical exponents
is performed for systems of linear size $L=512$. 
Numerical results for shallow quenches are presented in 
Section VI. In Section VII we make some concluding remarks.

\section{The hydrodynamic equations of motion}
\label{2.0}

There are two major 
differences between the hydrodynamics of simple liquids and 
that of liquid crystals.  First, the rod-like shape of the 
molecules means that they are rotated by gradients in the 
velocity.  Second, the equilibrium free energy is more complex 
than for a simple fluid and this in turn increases the 
complexity of the stress tensor in the
Navier-Stokes equation for the evolution of the fluid 
momentum. 
In a general formulation of liquid crystal hydrodynamics\cite{BE94},
the continuum equations of motion are written
in terms of a  tensor order parameter ${\bf Q}$ which is traceless 
and symmetric and is 
related to the direction of individual molecules $\vec {\hat{m}}$ by 
$Q_{\alpha\beta}= \langle \hat{m}_\alpha \hat{m}_\beta -
{1\over 3} \delta_{\alpha\beta}\rangle$  where the angular brackets
denote a coarse-grained average.
(We shall use Greek indices to represent Cartesian directions and
assume the usual  sum over
repeated indices.)
The advantage of this
approach is that it includes both the isotropic (${\bf Q}=0$) and the
nematic (${\bf Q} \neq 0$) phases and allows an order parameter of
variable magnitude within the latter. Hence it is possible to explore
the effect of flow on the phase transition between the two states.
Moreover the hydrodynamics of topological defects (point defects in
two dimensions) is naturally included in the equations.
We will study a two-dimensional system with the flow confined to the
$xy$-plane.  However we will allow the director field to point in any
direction ($x$, $y$ and $z$). As such the tensor order parameter is
always a $3x3$ matrix. 

The equilibrium properties of the liquid 
crystal are described 
 by the Landau-de Gennes free energy functional\cite{DG93}
\begin{equation}
{\cal F}=\int d^3 r \left\{ \frac{1}{2} (1-\frac{\gamma}{3})
  Q_{\alpha \beta}^2 - \frac{\gamma}{3} Q_{\alpha \beta}
Q_{\beta \gamma}Q_{\gamma \alpha}+ \frac{\gamma}{4}
  (Q_{\alpha \beta}^2)^2 + \frac{\kappa}{2} (\partial_\alpha Q_{\beta \lambda})^2
  \right\}.
\label{free}
\end{equation}
This free energy describes a first-order transition from the isotropic
to the nematic phase.
In general three elastic constants are needed to fully
characterize the nematic phase\cite{DG93} but we restrict 
ourselves to a single elastic constant $\kappa$. This can be shown to
be equivalent to having three equal Frank elastic constants when the
order parameter is uniaxial \cite{S74}.  This simplification
is not expected to affect the scaling behavior.

The order parameter ${\bf Q}$ is not conserved. It evolves according
to a convection-diffusion equation\cite{BE94,OD97,D81,FC98}
\begin{equation}
(\partial_t+{\vec u}\cdot{\bf \nabla}){\bf Q}-{\bf S}({\bf W},{\bf
  Q})= \bar{\Gamma} {\bf H}
\label{Qevolution}
\end{equation}
where ${\vec u}$ is the bulk fluid velocity and $\bar{\Gamma}$ 
is a collective rotational diffusion constant. A generalized form
of $\bar{\Gamma}$ is
\begin{equation}
    {\bar{\Gamma}} = \frac{\Gamma}{(1-{3 \over 2}Tr {\bf Q}^2)^2}
 \label{Gammanorm}
\end{equation}
where the ${\bf Q}$-dependence in the denominator enhances 
reorientation for well-ordered systems\cite{D81}. Note that in 
previous studies of the kinetics of phase separation for liquid crystals the 
${\bf Q}$-dependence in the diffusion parameter has been always 
neglected and $\bar{\Gamma} = \Gamma$ assumed. 

The first term on the left-hand side of equation (\ref{Qevolution})
is the material derivative describing the usual time-dependence of a
quantity advected by a fluid with velocity ${\vec u}$. This is
generalized by a second term 
\begin{equation}
{\bf S}({\bf W},{\bf Q})
=({\bf D}+{\bf \Omega})({\bf Q}+{\bf I}/3)+({\bf Q}+
{\bf I}/3)({\bf D}-{\bf \Omega})
-2({\bf Q}+{\bf I}/3){\mbox{Tr}}({\bf Q}{\bf W})
\end{equation}
where ${\bf D}=({\bf W}+{\bf W}^T)/2$ and
${\bf \Omega}=({\bf W}-{\bf W}^T)/2$
are the symmetric part and the anti-symmetric part respectively of the
velocity gradient tensor $W_{\alpha\beta}=\partial_\beta u_\alpha$.
${\bf S}({\bf W},{\bf Q})$  appears in the equations of motion because
the order parameter distribution can be both rotated and stretched by
flow gradients. 

The term on the right-hand side of equation (\ref{Qevolution})
describes the relaxation of the order parameter towards the minimum of
the free energy in a way analogous to Model A\cite{HH77}. 
The molecular field ${\bf H}$ which provides the driving
motion is related to the derivative of the free energy by
\begin{eqnarray}
{\bf H}&=& -{\delta {\cal F} \over \delta Q}+({\bf
    I}/3) {\mbox Tr}\left\{ \delta {\cal F} \over \delta Q \right\}.
\label{H(Q)}
\end{eqnarray}  \\
The flow of the liquid crystal of density $\rho$ obeys the continuity
\begin{equation}
\partial_t \rho + \partial_\alpha \rho u_{\alpha} =0
\label{continuity}
\end{equation}
and the Navier-Stokes equation 
\begin{equation}
 \rho\partial_t u_\alpha+\rho u_\beta \partial_\beta
u_\alpha=\partial_\beta \tau_{\alpha\beta}+\partial_\beta
\sigma_{\alpha\beta}+{\rho \tau_f \over
3}\partial_\beta((1-3\partial_\rho P_{0})\delta_{\alpha \beta}\partial_\gamma u_\gamma+\partial_\alpha
u_\beta+\partial_\beta u_\alpha)
\label{NS}
\end{equation}
where  $\tau_f$ is related to the viscosity and $P_0$ is the pressure,

\begin{equation}
P_0=\rho T-\frac{\kappa}{2} (\nabla {\bf Q})^2.
\end{equation}

The details of the stress tensor reflect the additional
complications of liquid crystal hydrodynamics with respect to simple
fluids. 
There is a symmetric contribution
\begin{eqnarray}
\sigma_{\alpha\beta} &=&-P_0 \delta_{\alpha \beta}
-3 H_{\alpha\beta}-\partial_\beta Q_{\gamma\nu} {\delta
{\cal F}\over \delta\partial_\alpha Q_{\gamma\nu}}
\label{BEstress}
\end{eqnarray}
and an antisymmetric contribution
\begin{equation}
 \tau_{\alpha \beta} = Q_{\alpha \gamma} H_{\gamma \beta} -H_{\alpha
 \gamma}Q_{\gamma \beta} .
\label{as}
\end{equation}
For the symmetric contribution we are using the form derived by
Doi\cite{D81}.  This is only quantitatively correct in the vicinity of
the transition with the general form being slightly more complex 
\cite{BE94}.  We do not expect any qualitative differences to result
from this difference in the regime in which we operate in this paper.

\section{A lattice Boltzmann algorithm for liquid crystal
hydrodynamics}

We now define a lattice Boltzmann algorithm which solves the
hydrodynamic equations of motion of a liquid crystal 
(\ref{Qevolution}), (\ref{continuity}), and (\ref{NS}). This section
may safely be omitted by readers interested in the physical results
but not in the details of the simulations.

Lattice Boltzmann algorithms are defined in
terms of a set of continuous variables, usefully termed partial
distribution functions, which move on a lattice in discrete space and
time \cite{C98}. 
The simplest lattice Boltzmann algorithm, which describes the
Navier-Stokes equations of a simple fluid, is defined in terms of a
single set of partial distribution functions which sum on each site to
give the density. For liquid crystal hydrodynamics this must be
supplemented by a second set, which are tensor variables, and which
are related to the tensor order parameter ${\bf Q}$.

We define two distribution functions, the scalars $f_i (\vec{x})$ and
the symmetric traceless tensors ${\bf G}_i (\vec{x})$ on each lattice
site $\vec{x}$. Each $f_i$, ${\bf G}_i$ is associated with a lattice
vector ${\vec e}_i$. We choose a nine-velocity model on a square
lattice with velocity vectors ${\vec e}_i=(\pm 1,0),(0,\pm 1), (\pm 1,
\pm 1), (0,0)$. Physical variables are defined as moments of the
distribution function
\begin{equation}
\rho=\sum_i f_i, \qquad \rho u_\alpha = \sum_i f_i  e_{i\alpha},
\qquad {\bf Q} = \sum_i {\bf G}_i.
\label{eq1}
\end{equation} 

The distribution functions evolve in a time step $\Delta t$ according
to
\begin{equation}
f_i({\vec x}+{\vec e}_i \Delta t,t+\Delta t)-f_i({\vec x},t)=
\frac{\Delta t}{2} \left[{\cal C}_{fi}({\vec x},t,\left\{f_i
\right\})+ {\cal C}_{fi}({\vec x}+{\vec e}_i \Delta
t,t+\Delta
t,\left\{f_i^*\right\})\right]
\label{eq2}
\end{equation}
\begin{equation}
{\bf G}_i({\vec x}+{\vec e}_i \Delta t,t+\Delta t)-{\bf G}_i({\vec
x},t)=\frac{\Delta t}{2}\left[ {\cal C}_{{\bf G}i}({\vec
x},t,\left\{{\bf G}_i \right\})+
                {\cal C}_{{\bf G}i}({\vec x}+{\vec e}_i \Delta
                t,t+\Delta t,\left\{{\bf G}_i^*\right\})\right]
\label{eq3}
\end{equation}

The left-hand side of these equations represents free streaming with
velocity ${\vec e}_i$, while the right-hand side is a
collision step which allows the distribution to relax towards
equilibrium. $f_i^*$ and ${\bf G}_i^*$ are first order approximations
to 
$f_i({\vec x}+{\vec e}_i \Delta t,t+\Delta t)$ and ${\bf G}_i({\vec x}+{\vec
e}_i \Delta t,t+\Delta t)$
respectively. They are obtained from equations
(\ref{eq2}) and (\ref{eq3}) but with $f_i^*$ and ${\bf G}_i^*$ set to
$f_i$ and ${\bf G}_i$.
Discretizing in this way, which is similar to a predictor-corrector 
scheme, has the advantages that lattice viscosity terms are eliminated
to second order and that the stability of the scheme is improved.

The collision operators are taken to have the form of a single
relaxation time Boltzmann equation, together with a forcing term
\begin{equation}
{\cal C}_{fi}({\vec x},t,\left\{f_i \right\})=
-\frac{1}{\tau_f}(f_i({\vec x},t)-f_i^{eq}({\vec x},t,\left\{f_i
\right\}))
+p_i({\vec x},t,\left\{f_i \right\}),
\label{eq4}
\end{equation}
\begin{equation} 
{\cal C}_{{\bf G}i}({\vec x},t,\left\{{\bf G}_i
\right\})=-\frac{1}{\tau_{\bf G}}({\bf G}_i({\vec x},t)-{\bf
G}_i^{eq}({\vec x},t,\left\{{\bf G}_i \right\}))
+{\bf M}_i({\vec x},t,\left\{{\bf G}_i \right\}).
\label{eq5}
\end{equation}

The form of the equations of motion and thermodynamic equilibrium
follow from the choice of the moments of the equilibrium distributions
$f^{eq}_i$ and ${\bf G}^{eq}_i$ and the driving terms $p_i$ and
${\bf M}_i$. $f_i^{eq}$ is constrained by
\begin{equation}
\sum_i f_i^{eq} = \rho,\qquad \sum_i f_i^{eq} e_{i \alpha} = \rho
u_{\alpha}, \qquad
\sum_i f_i^{eq} e_{i\alpha}e_{i\beta} = -\sigma_{\alpha\beta}+\rho
u_\alpha u_\beta
\label{eq6} 
\end{equation}
where the zeroth and first moments are chosen to impose conservation
of
mass and momentum. The second moment of $f^{eq}$ controls the symmetric
part of the stress tensor, whereas the moments of $p_i$
\begin{equation}
\sum_i p_i = 0, \quad \sum_i p_i e_{i\alpha} = \partial_\beta
\tau_{\alpha\beta},\quad \sum_i p_i
e_{i\alpha}e_{i\beta} = 0
\label{eq7}
\end{equation}
impose the antisymmetric part of the stress tensor.
For the equilibrium of the order parameter distribution we choose
\begin{equation}
\sum_i {\bf G}_i^{eq} = {\bf Q},\qquad \sum_i
{\bf G}_i^{eq} {e_{i\alpha}} = {\bf Q}{u_{\alpha}},
\qquad \sum_i {\bf G}_i^{eq}
e_{i\alpha}e_{i\beta} = {\bf Q} u_\alpha u_\beta .
\label{eq8}
\end{equation}
This ensures that the order parameter
is convected with the flow. Finally the evolution of the
order parameter is most conveniently modeled by choosing
\begin{equation}
\sum_i {\bf M}_i = {\bar \Gamma} {\bf H}({\bf Q})
+{\bf S}({\bf W},{\bf Q}) , \qquad
\qquad \sum_i {\bf M}_i {e_{i\alpha}} = (\sum_i {\bf M}_i)
{u_{\alpha}}.
\label{eq9}
\end{equation}
which ensures that the fluid minimizes its free energy at equilibrium.

Conditions (\ref{eq6})--(\ref{eq9})
can be satisfied as is usual in lattice Boltzmann
schemes by writing the equilibrium distribution functions and forcing
terms as polynomial expansions in the velocity\cite{C98}.
Taking the continuum limit of equations (\ref{eq2}) and (\ref{eq3}) and
performing a Chapman-Enskog expansion leads to the equations of motion
of liquid crystal hydrodynamics
(\ref{Qevolution}), (\ref{continuity}), and (\ref{NS})\cite{DO00}.

\section{Phase ordering kinetics}

\subsection{Measures}

The kinetics of phase separation in liquid crystals
has been examined using simulations by Zapotocky et al \cite{ZG95}
and more recently by Fukuda\cite{F98}. The former studied phase 
separation in the diffusive regime whilst the latter added hydrodynamics.
In both cases, rather deep quenches were performed and three different
quantities considered in order to make
a quantitative analysis of the domain growth. 

The first measure is the scalar correlation function for the
tensorial nematic order parameter ${\bf Q}$ defined by\cite{ZG95}
\begin{equation}
C({\bf r},t)=\frac{\langle {\rm Tr}[{\bf Q}({\bf 0},t){\bf Q}({\bf r},t)]\rangle}{\langle
  {\rm Tr}{\bf Q}^2({\bf 0},t)\rangle},
\end{equation}
where $\langle \cdots \rangle$ denotes averaging over the positions
${\bf 0}$. The correlation function is normalized so that 
$C({\bf 0},t)=1$. 
A correlation length $L_{cor}(t)$ at time t is defined by
\begin{equation}
C(L_{cor},t)=1/2.
\end{equation}

Dynamical scaling states that the system is dynamically self-similar
in time, except for a change in the length scale.
If dynamical scaling holds, the correlation length will control the
statistical properties of the system.  Plotting the correlation
function  as a
function of ${\bf r}/L_{cor}(t)$, the data at
different times should collapse onto a single curve. 
Moreover $L_{cor}(t)$ should decay with time as a power law
\begin{equation}
L_{cor}(t) \sim t^{\phi_{cor}}.
\end{equation}
Zapotocky et al.\onlinecite{ZG95} obtained an exponent $\phi_{cor}=0.41$
significantly lower than the value $1/2$ suggested by the diffusive 
character of the equation of motion for the order parameter
({\ref{Qevolution}) and by scaling arguments\onlinecite{L62}.

A second measure is the Fourier transform of the correlation function,
the structure factor $S(k,t)$.  The scaling form for
the structure factor in $d$-dimensions is
\begin{equation}
S(k,t)=L^d_{cor}(t)\,g(k L_{cor}(t))
\label{Sscales}
\end{equation}
where $g$ should have the form $g(y)\sim y^{-(N+d)}$ for large $y$ 
for the $O(N)$ vector model (Porod's law)\cite{B94}.  
To see how this arises note
that, for $y=k L_{cor} \gg 1$, the structure factor probes the order
parameter at length scales much smaller than the separation between
defects.  Substantial variation of
the order parameter on these length scales happens only in the
vicinity of the defect cores, and is not related to inter-defect
correlations.  This implies that
\begin{equation}
S(k,t)\sim \rho_{def}(t)b(k), \qquad\,\, k L_{cor} \gg 1,
\end{equation}
where $\rho_{def}$ is the density of defects in the system and $b(k)$
is a function of $k$ only.  If we further assume that the separation
of defects scales as the correlation length $L_{cor}$, so that
$\rho_{def}\propto L_{cor}(t)^{-(d-s)}$, where $s$ is the dimensionality
of the defect (zero here) then for $d=2$ 
scaling implies
\begin{equation}
g(y) \sim y^{-(d+(d-s))}=y^{-4}.
\end{equation}
Zapotocky and collaborators did find the $y^{-4}$ behavior in the
tails\cite{ZG95}.  They also noted that the scaling functions appeared to approach
the $y^{-4}$ law from above.
This effect has been observed experimentally
and it is attributed to inter-defect correlations.

A scaling form also holds for the elastic energy.  To see this
note from Eq. (\ref{free}) that
\begin{eqnarray}
{\cal F}_{el}& \propto &\int d{\bf r} (\partial_\alpha Q_{\beta\gamma})(\partial_\alpha
Q_{\beta\gamma})\nonumber\\
&=& \int d{\bf k} k^2 S(k,t)\nonumber\\
&=& \int d{\bf k} [k L_{cor}(t)]^2 g(k L_{cor}(t))
\end{eqnarray}
for $d=2$ where the last line has been obtained by using the scaling 
form Eq. (\ref{Sscales}).  The integral can be split into three
parts, $0 < r < a$, $a < r < L_{cor}(t)$ and $L_{cor}(t) < r <
\infty$, where $a$ is a cutoff.  Assuming that the result is not
dependent on the cutoff, the first part can be neglected, whereas for
the second contribution using Porod's law for $g$ and integrating from $a$
to $L_{cor}(t)$ gives 
\begin{equation}
{\cal F}_{el} \propto L_{cor}(t)^{-2}\ln[L_{cor}(t)/a].
\label{Fdscal}
\end{equation}
The last term can be neglected since it
should scale roughly as $L^{-2}_{cor}(t)$ assuming that 
$g(y)$ is approximately
constant at small $y$.
Differentiating Eq. (\ref{Fdscal})
\begin{equation}
\frac{d \ln {\cal F}_{el}}{d \ln t} = \frac{d \ln L_{cor}(t)}{d \ln t}\frac{d \ln {\cal F}_{el}}{d
\ln L_{cor}}
\sim  \phi_{cor}(t)\left(-2+\frac{1}{\ln[L_{cor}(t)/a]}\right)
\end{equation}
where $\phi_{cor}(t)$ is the exponent from the scaling of the
correlation length.  We define a new exponent for the characteristic
energy as
\begin{equation}
\phi_{el}(t)=-\frac{1}{2}\frac{d \ln {\cal F}_{el}}{d \ln t}\sim \phi_{cor}(t)-\frac{\phi_{cor}(t)}{2\ln[L_{cor}(t)/a]}.
\label{phid}
\end{equation}
Note the
logarithmic corrections to the scaling of
the elastic energy. Taking these into account, Zapotocky et
al. obtained reasonable agreement between $\phi_{el}$ and $\phi_{cor}$,
whereas without taking them into account they obtained
$\phi_{el}=0.325 < \phi_{cor}$ \cite{ZG95}.

A length scale can also be obtained from the average 
separation of topological defects.  This separation is found by
counting the number of defects in the system to obtain $\rho_{def}$
and then taking 
\begin{equation}
L_{def}(t)=1/\sqrt{\rho_{def}}.
\end{equation}
Zapotocky
et al. obtained a value  $\phi_{def}=0.374$
for the corresponding exponent.  It is assumed
that defect annihilation is the process which controls the growth of
the correlation length and therefore dynamical scaling implies
$\phi_{cor} = \phi_{def}$. 
Zapotocky
et al. could give no explanation of the discrepancy in their values
and interpreted it as  an indication of the 
violation of dynamical scaling.

Indeed in almost all of the  published numerical 
work on phase ordering in systems which
order via the annihilation of topological defects, it has been found that
the exponents are less than the value $1/2$ expected from dimensional
analysis.  A suggestive argument for the origin of this discrepancy 
has been given by Yurke et al \cite{YP93} for the $O(2)$ model.  These
authors
obtained an approximate equation of motion for an isolated
defect-antidefect pair by equating the attractive and frictional
forces acting on each defect.  The attractive force was assumed to have
the form $F_{at} \propto -1/L$ where $L$ is the separation of the
defects. This would seem to be a reasonable assumption as it is well 
known the energy of such a pair goes like $\ln L$ \cite{DG93}.  
The frictional force was taken to be $F_{fr}\propto v \ln (R/a)$ where
$R$ is the ``size'' of a defect, $a$ its  core size, and
$v=\frac{1}{2}\frac{dL}{dt}$ its velocity.  The ``size'' is then
assumed to be equal to $L$.  Assigning a defect size of $L$
may seem somewhat questionable. However the assumption can be
supported by a more rigorous argument \cite{CD96}.

Equating the frictional and elastic forces gives an implicit
formula for the defect separation as a function of the time $t$ 
before annihilation
\begin{equation}
L_{def}(t) \sim\left[\frac{t}{\ln[L_{def}(t)/a]-1/2}\right]^{1/2}.
\label{Ldef}
\end{equation}
The growth in this length scale for increasing times before annihilation
is then assumed to be the same as the growth in the average separation
of defects after a quench.  This gives an apparent exponent
\begin{equation}
\phi_{def} \propto \frac{1}{2} \left[ \frac{
    t}{L_{def}^2(t)\ln[L_{def}(t)/a]} \right].
\label{phiLdef}
\end{equation}
The argument implies that the failure to measure $t^{1/2}$ is
due to logarithmic corrections. However it has been claimed \cite{ZG95} 
that this is effectively untestable because of the unknown constant of
proportionality in Eq. (\ref{phiLdef}).  One
can however, measure an effective exponent for the two defect case and
compare to that for the average separation after a quench.  Zapotocky et
al obtained a similar value, $\phi_{def}=0.375$\cite{ZG95}.  

More recently Fukuda\cite{F98} has investigated the effect of the
stress-induced flow
on the kinetics of the nematic-isotropic transition
by numerically solving the hydrodynamic equations for the 
tensor order parameter {\bf Q} and the fluid velocity in $d=2$.
The equations of motion were slightly different
from the ones considered here. 
An older model which considers a corotational derivative
rather than an upper convective derivative\cite{OG92} was used.   
For the purely
dissipative case Fukuda obtained similar exponents to Zapotocky et al; 
for the hydrodynamic case he obtained  
$\phi_{cor}\simeq \phi_{def} = 0.43$. Note that, in contrast to the 
case without flow,
$\phi_{cor}\simeq \phi_{def}$
indicating that the dynamical scaling hypothesis is confirmed when
hydrodynamics is included. However the value calculated is 
still lower than the value $1/2$ expected for these systems.

Most experiments on phase ordering in liquid crystals are in the three
dimensional regime where the defect line energy controls the dynamics.
There are only a limited number of experiments in a
two dimensional geometry \cite{WWLY93,PGY94}.  Even in the
experiments it is difficult to reach the asymptotic regime
and the expected power law of $t^{-1}$ for the number of defects is
only obtained for the last decade of time \cite{PGY94}.

\subsection{Corrections to Scaling and Parameter space}

Our goal is to ascertain why previous studies have not measured
the expected exponents and then, if possible, to perform simulations
in the asymptotic scaling regime to obtain precise exponent values.
In order to do this we first determine numerically the dependence of
the corrections to scaling on the parameters at our disposal.  In
doing this we will see that the difficulty in obtaining
the asymptotic values of the exponents may have had as much to do with the
fine scale details of the simulation on the lattice as with the
constraints of lattice size and run times.

As it seems clear from all previous studies that the phase behavior 
is dominated by the motion of ${\bf Q}$ in the plane, we restrict 
the order parameter to be of the form
\begin{equation}
${\bf Q}$=\left(
\begin{array}{ccc}
Q_{xx} & Q_{xy} & 0\\
Q_{xy} & Q_{yy} & 0\\
0 & 0 & -(Q_{xx}+Q_{yy})
\end{array}
\right),
\end{equation}
in order to speed up the numerics.
The quenches that we consider are envisaged to start at a high 
temperature, at which the equilibrium phase is the disordered, 
isotropic phase. We take as the initial condition for the lattice Boltzmann 
simulations a configuration representative of this phase in which the
$Q_{\alpha \beta}$ at each lattice point are random numbers uniformly
distributed in a small interval $[-\delta,\delta]$.  More precisely,
by writing
\begin{eqnarray}
    Q_{xx} &=& A(3\cos\phi\cos\phi-1)/2, \nonumber\\
    Q_{xy} &=& 3 A \cos\phi\sin\phi/2, \nonumber\\
    Q_{yy} &=& A(3\sin\phi\sin\phi-1)/2
\end{eqnarray}
the initial configuration is obtained by assigning, at each 
lattice point, a random number between $[-0.02,0.02]$ for the 
amplitude $A$ and a random number between $[-\pi,\pi]$ for the
phase $\phi$.

To ascertain a rough dependence of the corrections to scaling on
the various parameters of the system we
examined quenches below and close to the spinodal line
for several sets of parameters values.  
For this exploration of parameter space we used a lattice size of $256
\times 256$ and 
averaged over only two to five runs. 

Figure~\ref{sch} shows the Schlieren pattern and associated fluid
flow from a typical run.  The shading indicates the orientation of the
director field and the arrows the fluid velocity.  The thing to note
in such a diagram is the intersection of light and dark ``brushes''
which indicate the location of the $\pm 1/2$ disclinations. It is
apparent that
the main features of the flow are the vortices
associated with the moving disclinations.  The interaction of vortices
in two dimensional flow has the same form as the interaction of
disclinations in the director field so one does not expect any
qualitative changes in the scaling behavior.

We now compare 5 sets of data. The first four data sets are a quench to
$\gamma=3.25$ (note that the first order transition is at $\gamma=2.7$ and 
that the spinodal is at $\gamma=3.0$).  The first is a baseline run. In the
second ${\bar{\Gamma}}$ is given by Eq.(\ref{Gammanorm}) whereas in
the others ${\bar{\Gamma}}$ is taken to be a constant and equal to
$1$. In the third we turn off hydrodynamics by relaxing the momentum
conservation constraint (i.e. the second Eq. in (\ref{eq6})) at time
$t=5000$. The fourth has a value of $\kappa$ ten times smaller
than the rest.  The fifth data set is similar to the first except now
the quench is deeper, to $\gamma=3.7$. 

The correlation function for  $\gamma=3.25$ and with the
renormalized diffusion constant is shown in Figure~\ref{correlat}.
There is a reasonable collapse of the data when the correlation
function is plotted against $r/L_{cor}(t)$.
This is as expected from dynamical scaling.
However even if a  function does appear to collapse it is very hard to
measure the quality of the collapse and hence this provides only a
weak test of scaling. 

Figure \ref{Lcor} shows the length scale obtained from the correlation
function $L_{cor}$ as a function of time for 
all five data sets.  The exponents $\phi_{cor},\phi_{def},\phi_{el}$ 
obtained by fitting this data, as well as similar data for the
number of defects and the distortion energy, to power laws are given
in Table \ref{exponents}. 
Although, as there are corrections to scaling, this is not a good way
of obtaining the exponents, it does allow for a clear comparison
between the different parameter sets. 
A more sophisticated way of estimating the growth exponents 
is to obtain an
effective exponent $\phi(t^{*})$
by performing a simple linear fit to the plots of $\ln L$ versus $\ln
t$ (e.g. Figure~\ref{Lcor})
for data in the interval  $t_{-m},t_{-m+1}\cdots, t^{*},\cdots t_{m-1},t_{m}$.
$m=8$ is used in the fits presented here.

From these preliminary runs with $L=256$ we can make the following 
observations:

\begin{itemize}
    
\item A renormalized diffusion constant $\bar{\Gamma}$ seems to help
  in the sense that the exponents are closer to the expected
  asymptotic value of 1/2.  However this turns out to be mostly an
  early time effect.  To see this we plot the effective correlation length
  exponent  as a function of time in Figure~\ref{phicor} for the first
  two data sets in Table~\ref{exponents}.  We see that for the
  baseline data set, the exponent first decreases and then increases
  with time whereas with the renormalized diffusion constant
  $\bar{\Gamma}$ the exponent immediately starts increasing with time
  toward the asymptotic value. Thus, we are effectively accessing
  longer times with the renormalized diffusion constant. This is
  reasonable because the initial ordering in the system, where the
  order parameter saturates away from the defect cores, is a diffusive process.

\item Hydrodynamics appears to speed the ordering very slightly, but
  not by a significant amount.  The crucial test here is the run
  denoted in Figure~\ref{Lcor} by filled triangles where at a given time step
  (t=5000) we have explicitly switched off the hydrodynamics by
  randomizing the velocities directions after each collision step.
  One can see in Figure~\ref{Lcor} that no significant change occurs
  with respect to the run in which the hydrodynamics is fully taken
  into account.  This conclusion agrees with previous results by
  Fukuda\cite{F98} even though his formulation of the model is somewhat
  different from ours. 

\item Reducing $\kappa$ reduces the size of the defect cores and we
  find that changing $\kappa$ has a significant effect on the
  results. To see this we compare results for the baseline run with
  $\kappa=0.02$ to the
run with a value of $\kappa$
  ten times smaller ($\kappa=0.002$). The effective exponent extracted
  from
  the correlation function is shown in Figure~\ref{phicor2}.  There
  are a few important things to note  from this Figure.  First
  the effective exponent is not a monotonic function of time:  it
  reaches a maximum value at a given time and then it decreases as
  time is  further increased. This is true  for the baseline case as well,
  but  the time at which it happens  is much larger than for the run
  with more localized defects.
Note that this effect has also been
  observed in previous simulations \cite{ZG95} and interpreted as the
  onset of freezing in the dynamics, due mainly to the finite size of
  the system simulated. However
the fact that the effect
  is stronger for smaller values of $\kappa$ suggests instead a lattice
  discretization effect.

As an obvious length scale in this
  problem is the size of a defect core $a$ it is
  tempting to take $\kappa$ as small as possible to minimize this scale.
However,
  once a defect core is of the order of a lattice spacing, it becomes
  energetically advantageous for a defect to be centered in a
  plaquette of the lattice as this will minimize the distortion in the
  amplitude of the order parameter  which is actually realized on the lattice,
  as shown in Figure~\ref{defcut}.  
  This creates an additional potential which can trap the defect.  The
  height of this potential is a decreasing function of $\kappa$
  as, if the amplitude distortion caused by a defect is spread over many
  sites, the defect will be less sensitive
  to exactly where it is centered. 

 This effect occurs at
  later times in the simulation because
inter-defect interactions decrease as the defect spacing
increases.   At the point where  the lattice pinning potential balances
  the defect-defect interactions motion will stop. Indeed even 
before this time, the lattice pinning will cause an
  additional frictional drag on the defects which could be detrimental
  in trying to fit to theory.  Thus we see that taking $\kappa$ small
  to try to achieve longer length scales (as appears to have been done
  in almost all previous work) is a trap better avoided.  

  Second, as can be seen from Figure~\ref{phicor2} , if the time is scaled by
  $\kappa$ the effective exponent for the
  baseline case appears to continue on the same curve as for smaller
  $\kappa$.  At first this seems surprising since
  one might expect time to scale with the diffusion constant, but not
  necessarily with $\kappa$.  To see why it occurs one can
  substitute a form for the director of a uniaxial nematic away from 
the defect
  core, ${\bf n}=(\cos \theta,\sin \theta)$ into Equation
  (\ref{Qevolution}).  Ignoring the hydrodynamic flow one obtains a
  diffusion equation for~$\theta$ 
  \begin{equation}
    \partial_t \theta= D_\theta \nabla^2 \theta,
\label{diffusion}
  \end{equation}
  where $D_\theta=\kappa \bar{\Gamma}$, confirming that the relevant diffusion
  constant is proportional to both  $\bar{\Gamma}$ and 
$\kappa$ \cite{us00,CD96}.
  
\item Finally, the data for deeper quenches show an exponent further
  from $1/2$.
This is also due to the pinning potential described
  above.  As can be seen from the effective exponents shown
  in Figure \ref{phicor3} the freezing happens at an earlier time for
  the deeper quench.  This is not surprising as defects are more localized
further from the transition and  hence the
  lattice pinning potential will be greater.
\end{itemize}

\section{Asymptotic results: critical exponents}

In the previous section we examined the kinetics of phase ordering
in a system of linear size $L=256$ averaging over only a few initial 
configurations. This allowed us to explore in a more systematic way
the space of the  parameters (i.e. $\gamma$, $\kappa$, 
$\bar{\Gamma}$) relevant to the kinetics. 
In particular we found that by using a renormalized diffusion 
constant $\bar{\Gamma}$ and  values of $\kappa$ big enough to avoid 
pinning effects better scaling behavior  is achieved in less
time. Using this knowledge we now 
simulate bigger systems and
average over more starting configurations to obtain more precise
results for
the dynamical critical exponents of the model. 

We focus on the scaling behavior of the 
correlation function, the average defect separation and the elastic 
energy for phase separating systems of linear size $L=512$. Averages
were taken over $20$ initial configurations.
Quenches were run for $\gamma = 3.25$ and $\gamma = 3.0$. 
In both cases $\kappa = 0.3$ and 
$\bar{\Gamma} = 1/(1-{3 \over 2}Tr {\bf Q}^2)^2$. 

Figure \ref{Lcor512} shows the time evolution of the length scale
obtained from the correlation function for the two data sets.
Note that at late times the data (on the log-log plot)
displays a linear behavior with slope close to $1/2$.
A simple linear fit
of the data with $t \ge t_{x}=2239$ gives
\begin{equation}
    \phi_{cor} = 0.50 \pm 0.01
\end{equation}
for $\gamma = 3.25$, and
\begin{equation}
    \phi_{cor} = 0.48 \pm 0.02
\end{equation}
for $\gamma = 3.00$.  The errors were obtained by varying $t_{x}$
between $502$ and $5012$.  This systematic error
was much more significant than statistical errors evaluated from the
$\chi^{2}$ statistics of the linear fit.

As we have argued in the previous section a better quantity to check 
is the effective exponent $\phi(t^{\ast})$ as a function of fitting 
time $t^{\ast}$.
In Figure~\ref{phicor512} we show the effective exponent for the
correlation length  $\phi_{cor}(t^{\ast})$ as a 
function of inverse time.  
Note that the asymptotic value of $1/2$ is clearly reached for the 
last decade in time. As far as we know this is the first time that 
 strong evidence   has been obtained for the conjectured $1/2$
dynamical exponent for a
phase separating nematic liquid crystal.

Similar plots for the effective exponent for the number of defects 
$\phi_{def}(t^{\ast})$ are shown
in Figure~\ref{phidef512}. By looking at the high $t^{\ast}$ behavior
we obtain

\begin{equation}
\phi_{def} = 0.47 \pm 0.03 \quad \hbox{for}\quad \gamma = 3.25,
\qquad \phi_{def} = 0.48 \pm 0.03 \quad
\hbox{for}\quad \gamma = 3.00
\end{equation}
where the errors have been estimated from the fluctuations
in the effective exponents from a moving average with fewer points
( 4 in this case).
Again, we see that the
asymptotic value of the exponent is consistent with $1/2$.

Finally Figure~\ref{phien512}(a) shows the exponent obtained from the
elastic energy $\phi_{el}(t^{\ast})$ .  In this case a simple linear fit produces an 
asymptotic value smaller than $1/2$ ($\sim 0.4$ in the plot). This is 
mainly due to the logarithmic corrections to the scaling of the elastic 
energy (Eq.~\ref{Fdscal}) 
that depress the effective growth exponent for the characteristic 
energy length with respect to the correlation length exponent 
(Eq.~\ref{phid}). Indeed by adding the logarithmic corrections to
scaling as in Equation~\ref{phid}, we can bring the exponent into
agreement with that obtained from the correlation function, as shown
in  Figure~\ref{phien512}(b). 

By extrapolating the effective exponents at high $t^{\ast}$ values
we obtain:  
\begin{equation}
    \phi_{el} = 0.47 \pm 0.03
\quad \hbox{for} \quad \gamma = 3.25,\qquad \phi_{el} = 0.50 \pm 0.03 \quad
\hbox{for} \quad \gamma = 3.00.
\end{equation}

\section{Shallow quenches}
So far, we have examined the late stage kinetics of the orientational
ordering, long after the amplitude has ordered.  In this regime, the
ordering has shown dynamics consistent with that expected for a
diffusive XY model.  It is also interesting to examine the early stage
dynamics and shallower quenches where one may observe an interaction
between the two symmetries of the problem (i.e. those of the scalar
order parameter and the essentially $O(2)$ symmetry of the
orientational order).  

First we examine what is naively expected for the amplitude 
ordering dynamics.  Assuming that the orientational
degrees of freedom are  completely ordered
the free energy can be written in terms of just the amplitude of the
order parameter $q$\cite{D81}
\begin{equation}
{\cal F}={1 \over 2}(1-{\gamma \over 3})q^2-{\gamma \over 9} q^3 +
{\gamma \over 6} q^4.
\end{equation}
This is shown as the solid curves in Figure~\ref{freenwor} for the
cases $\gamma=2.7$ (phase transition point) and $\gamma=3.0$
(spinodal line, where the barrier between the isotropic and nematic
states vanishes).  One expects the presence of a free energy
barrier between the isotropic and nematic states to affect the phase
ordering dynamics.  In the presence of a barrier a finite
region of the nematic phase must be nucleated to initiate growth.
Beyond the spinodal 
(which occurs at $\gamma=3.0$) the system will order
spontaneously by spinodal decomposition throughout space.
If we artificially orient the director field before quenching, this is
indeed what is observed.  

However, in a normal quench where the
director is not oriented, one observes somewhat different behavior.
In particular, there is a region where one might expect spontaneous
ordering ($3.0<\gamma<\gamma_s(\kappa,L)$) which in fact exhibits a two stage
growth process in the absence of a nucleating site, and domain growth
from a nucleating site (as opposed to spinodal decomposition) if one
is provided.

Figure~\ref{ampvst} shows the amplitude as a function of time 
for $\gamma=3.05$ for
a system
which is initially disordered and in which we have not supplied any
strong nucleating sites.
There are clearly  two different stages of growth.  In
the first stage the amplitude grows very slowly for a long
time.  Then the system abruptly  orders, very quickly reaching the
equilibrium value.  If instead we supply the system with nucleating
sites at the beginning of the simulation it orders much more quickly
with domains growing from the nucleating sites.
These observations are consistent with the presence of a free energy barrier
which decays with time. 
 
The free energy barrier can be identified as resulting from the free energy
of the orientational order.  There is a time dependent quadratic term
in the free energy due to the distortion.  The distortion energy is
proportional to $\kappa q^2$ (see Eq. (\ref{free})) so that one has
an effective term
\begin{equation}
{\cal F}_d \propto \kappa q^2  L_{def}(t)^{-2}\ln[L_{def}(t)/a],
\label{extraF}
\end{equation}
using Eq. (\ref{Fdscal}) with  $L_{def}(t)$ being the separation
between defects.  As a result, not only does the state at $\gamma=3.0$
remain metastable but the global free energy minimum at $q > 0$ may not
be a minimum at zero time.  This contribution to the free energy is
included  in
Figure~\ref{freenwor} for different values of $L_{def}$.

The domain growth in the nucleated case is now easily understood.  Why
there is eventually ordering with no nucleation sites still requires
some explanation.  Eq. (\ref{extraF}) gives 
the average free energy contribution from the orientational order.
However, locally there are regions where there is higher orientational
order so that some gain in free energy can be made by increasing $q$
in these regions.  These regions are still not large enough
to gain sufficiently from this to pay for the cost of an interface between
fully ordered and disordered states so the elastic energy suppresses
the growth in $q$, keeping its typical value very small.  However, as
we have already pointed out in Eq. (\ref{diffusion}) the diffusion
constant for the orientational ordering is independent of $q$, as long
as $q>0$.  Hence the orientational ordering can proceed. (See, for
instance, the number of defects as a function of time shown in
Figure~\ref{Ndvst}.)  As the
orientation proceeds, $L_{def}$ grows and eventually the free energy barrier
is small enough that the system can form a sufficiently 
large region of the amplitude
ordered phase to nucleate growth.   Once such a region is formed 
the amplitude  ordering can proceed rapidly.

\section{Summary and Discussion}

To summarize: we have used a lattice Boltzmann algorithm to
investigate the phase ordering of liquid crystals quenched from the
isotropic to the nematic phase. The system is described by the
Beris-Edwards equations of liquid crystal hydrodynamics. These are
written in terms of a tensor order parameter which means that the
dynamics of topological defects which drives the phase ordering
appears naturally in the simulations. The liquid crystal moves towards
an equilibrium described by the Landau-de Gennes free energy.

The flexibility of the numerical scheme allows an investigation into
how the ordering process is affected by the model parameters. In
particular we find that if the length scale defined by the size of the
topological defects is taken to be too short the cores are pinned by
the lattice as the simulation proceeds giving an incorrect value for
the growth exponent. This is likely to have been the problem faced by
other authors who consistently obtained a value less than the $1/2$
expected from scaling arguments.

Averaging over 20 runs on a lattice of linear size 512 we obtained
$\phi_{cor} = 0.49 \pm 0.02$, $\phi_{el} = 0.485 \pm 0.03$, $\phi_{def} = 0.475 \pm 0.03$.
The value for $\phi_{el}$ was fit using logarithmic
corrections. Hydrodynamic flow speeds up the ordering slightly but
does not change the scaling behavior. This is reasonable because, in
two dimensions, interactions between the flow vortices set up by the
moving defects have the same logarithmic form as the defect-defect
interactions themselves. Thus theoretical and numerical
results are now in pleasing agreement.

\newpage

\begin{table}
\begin{tabular}{llccc}
{\it run} &{\it parameter set} & \multicolumn{3}{c} {\it exponents}\\  
              &    & $\phi_{cor}$ &  $\phi_{def}$  & $\phi_d$    \\ 
\hline
baseline&$\gamma=3.25,\bar{\Gamma} = 1, \kappa = 0.02$      & $0.42$    & $0.406$ & $0.35$   \\
renorm. diffusion&$\gamma=3.25, \bar{\Gamma}=(1-{3 \over 2}Tr {\bf Q}^2)^{-2}, \kappa = 0.02$  & $0.45$    & $0.43$  & $0.365$  \\
no hydrodynamics&$\gamma=3.25, \bar{\Gamma}= 1 \kappa = 0.02$ & $0.42$    & $0.398$ & $0.345$  \\
more localized defects&$\gamma=3.25, \bar{\Gamma} =1, \kappa =0.002$   & $0.36$    & $0.33$  & $0.27$   \\
deeper quench&$\gamma=3.7,  \bar{\Gamma} =1,\kappa  =0.02$     & $0.39$    & $0.36$  & $0.31$   \\
\end{tabular} 
\caption{Comparison of the exponents obtained by fitting the data from runs on small
  systems (e.g. Fig.~\ref{phicor} 
to a power law.  The statistical
  errors $\sim$ $0.003$ and the systematic errors $\sim$
  $0.015$ (as the end-points of the fit are changed).}
\label{exponents}
\end{table}

\newpage
\begin{figure}
\centerline{\epsfxsize=5.in
\epsffile{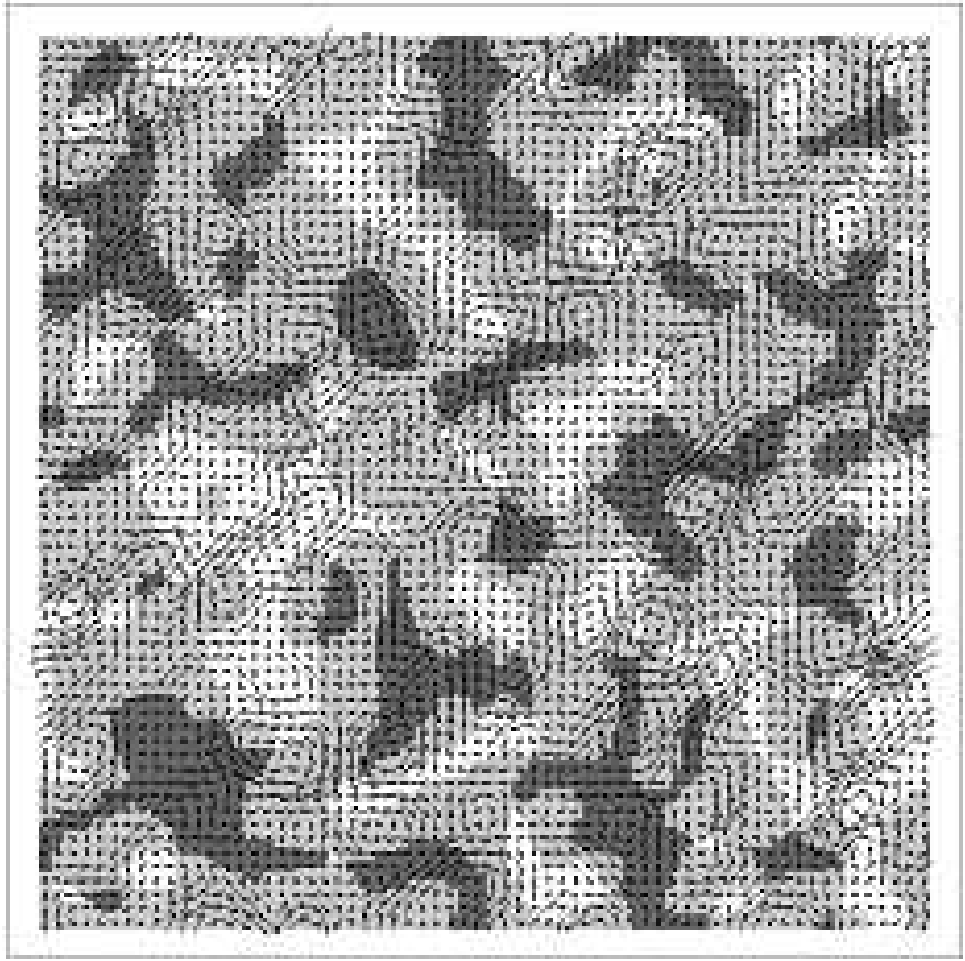}}
\vskip 0.2true cm
\caption{Schlieren pattern (shading) and fluid velocity (vectors)
associated with the  ordering of a liquid crystal after a quench from
the isotropic to the nematic phase.  The director in the darkest
regions is perpendicular to the director in the lightest regions (with
the other shade in between).  The shading on the Schlieren
pattern has been rendered with only three shades of gray so as not to
obscure the vector field.  Disclinations of strength $\pm 1/2$ are 
located at the intersection of dark grey and white ``brushes''.}
\label{sch}
\end{figure}

\newpage
\begin{figure}
\centerline{\epsfxsize=5.in
\epsffile{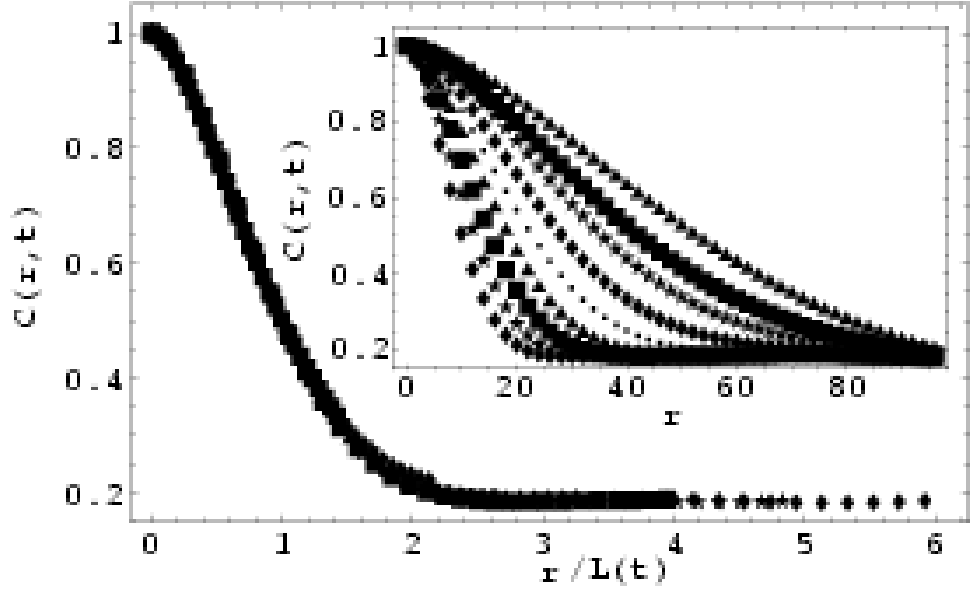}}
\vskip 0.3true cm
\caption{Correlation function for $\gamma=3.25$ and with renormalized
  diffusion constant.  Inset: unscaled correlation function for times
  between $t=159$ and $8913$.  Main figure:  correlation function as a
  function of $r/L(t)$ where $L(t)$ is determined as the length at
  which the unscaled correlation function crosses $1/2$.}
\label{correlat}
\end{figure}

\begin{figure}
\centerline{\epsfxsize=5.in
\epsffile{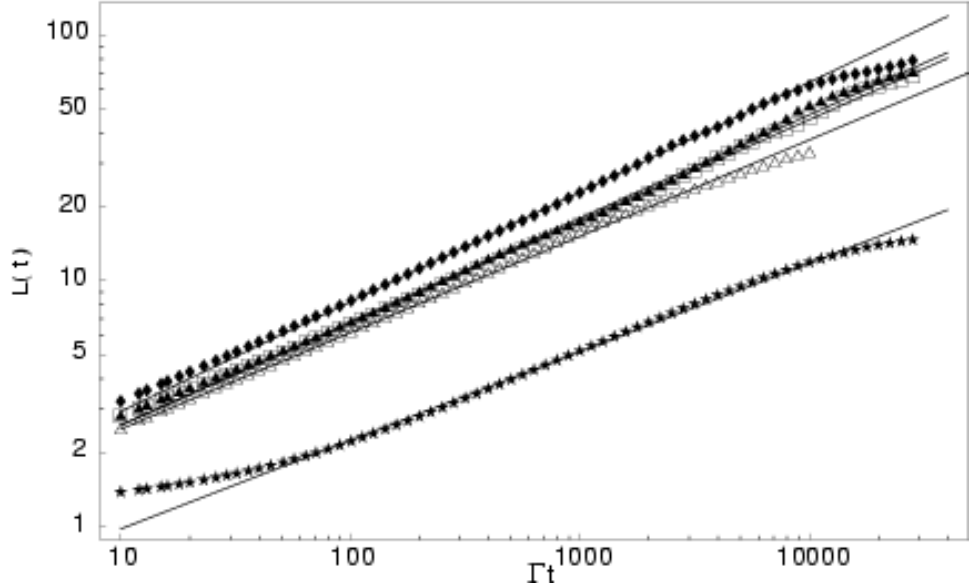}}
\vskip 0.2true cm
\caption{Length scale obtained from scaling of correlation function as
  a function of time.  The lines are power law fits and the symbols
  correspond to box: $\gamma=3.25$ baseline, diamond: $\gamma=3.25$
  renormalized 
  diffusion constant, filled triangle: $\gamma=3.25$ hydrodynamics off,
  star: $\gamma=3.25$ small kappa, hollow triangle: $\gamma=3.7$ as
  baseline.   In this figure (and all others) times are multiplied by
  $\Gamma$ to make them dimensionless and distances are measured in
  units of the lattice spacing of the simulation.}
\label{Lcor}
\end{figure}

\begin{figure}
\centerline{\epsfxsize=5.in
\epsffile{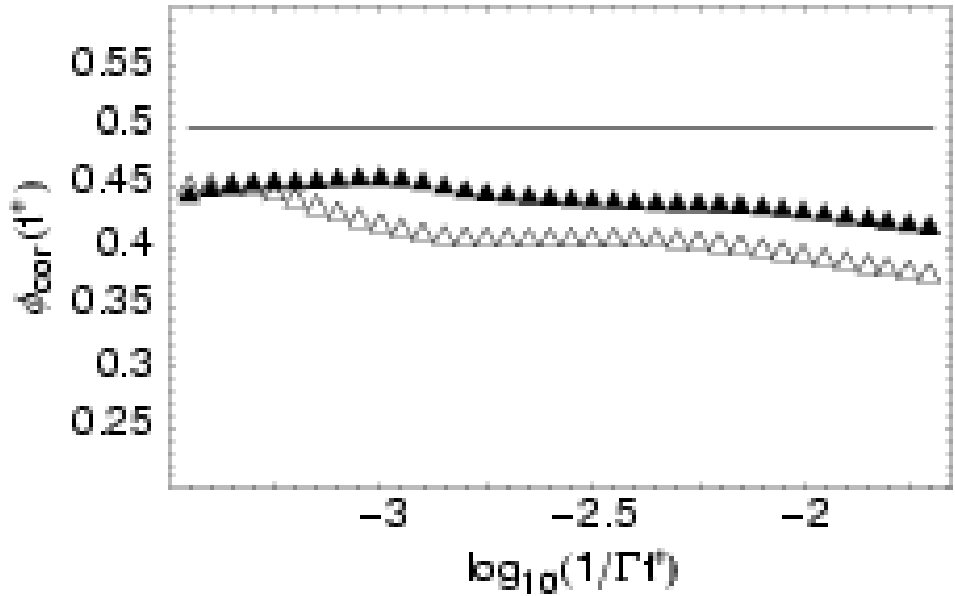}}
\vskip 0.2true cm
\caption{Effective exponent for the growth of the correlation length
 as a function of time empty triangle: $\gamma=3.25$ baseline, filled triangle :
  $\gamma=3.25$ renormalized diffusion constant.}
\label{phicor}
\end{figure}

\begin{figure}
\centerline{\epsfxsize=5.in
\epsffile{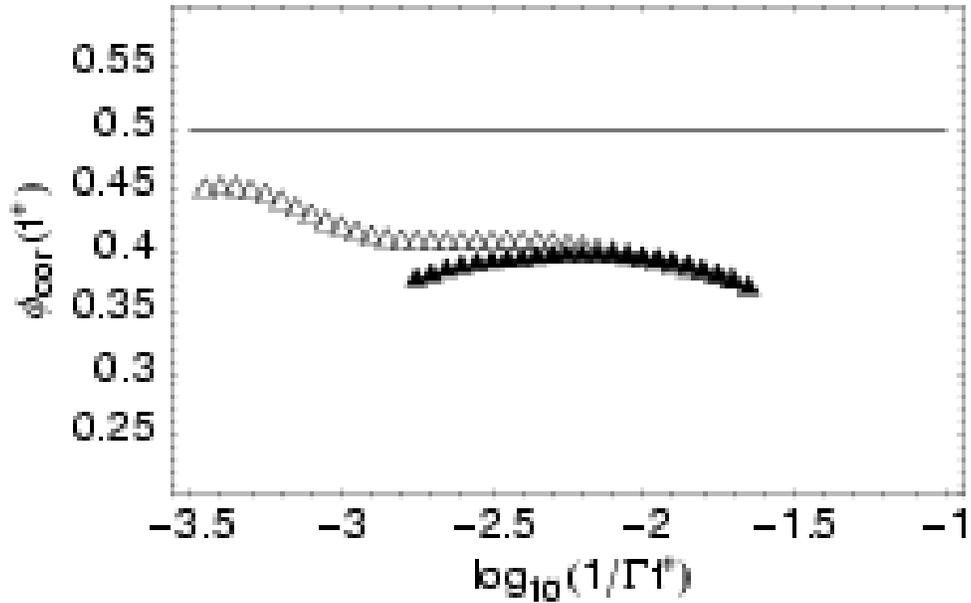}}
\vskip 0.2true cm
\caption{Effective exponent for the growth of the correlation length
  as a function of time empty triangle: $\gamma=3.25$ baseline, 
filled triangle: $\gamma=3.25$ $\kappa=1/10$ of the baseline value.
 The time $t^*$
 has been multiplied by $1/10$ (i.e. time is scaled by $1/\kappa$) for the
 small kappa case. }
\label{phicor2}
\end{figure}

\begin{figure}
\centerline{\epsfxsize=5.in
\epsffile{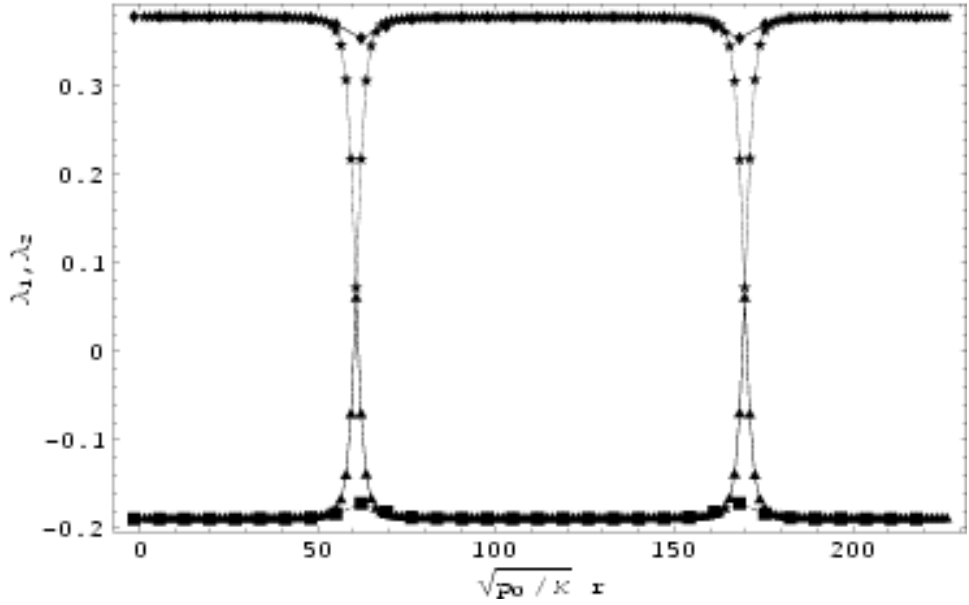}}
\vskip 0.2true cm
\caption{Two largest eigenvalues of ${\bf Q}$ along a cut joining the
  center of two $\pm 1/2$ defects moving to annihilate each other.  The
  squares and diamonds are for the case with $\kappa=0.02$ and the
  stars and triangles for $\kappa=0.5$.  The dips are an indication of
  the lattice pinning potential. The abscissa is scaled by the
  length $\sqrt{\kappa/p_0}$ to make it dimensionless (all other
  parameters are the same).}
\label{defcut}
\end{figure}

\begin{figure}
\centerline{\epsfxsize=5.in
\epsffile{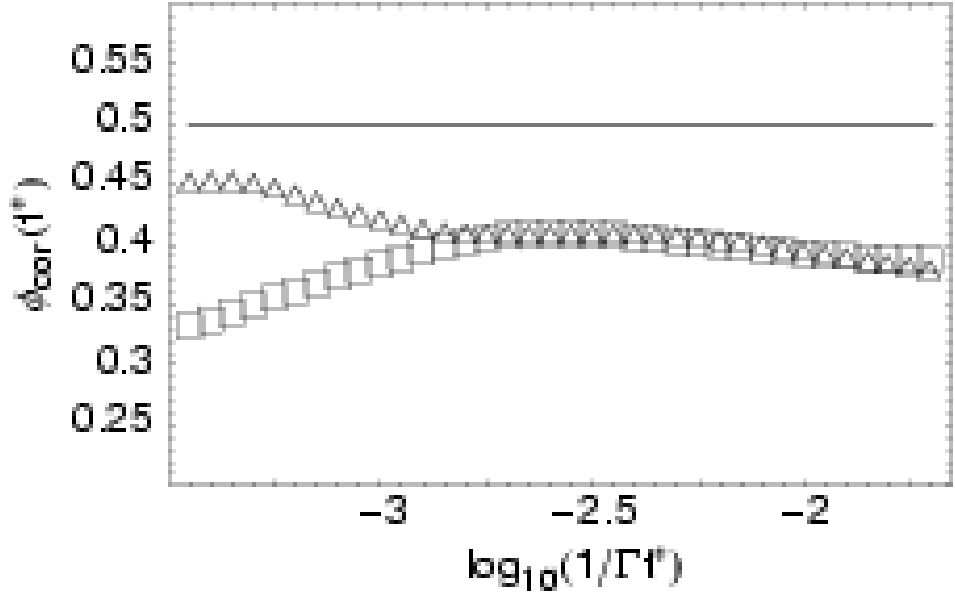}}
\vskip 0.2true cm
\caption{Effective exponent for the growth of the correlation length
  as a function of time  triangle: $\gamma=3.25$ baseline, box:
  $\gamma=3.7$ baseline.}
\label{phicor3}
\end{figure}

\begin{figure}
\centerline{\epsfxsize=5.in
\epsffile{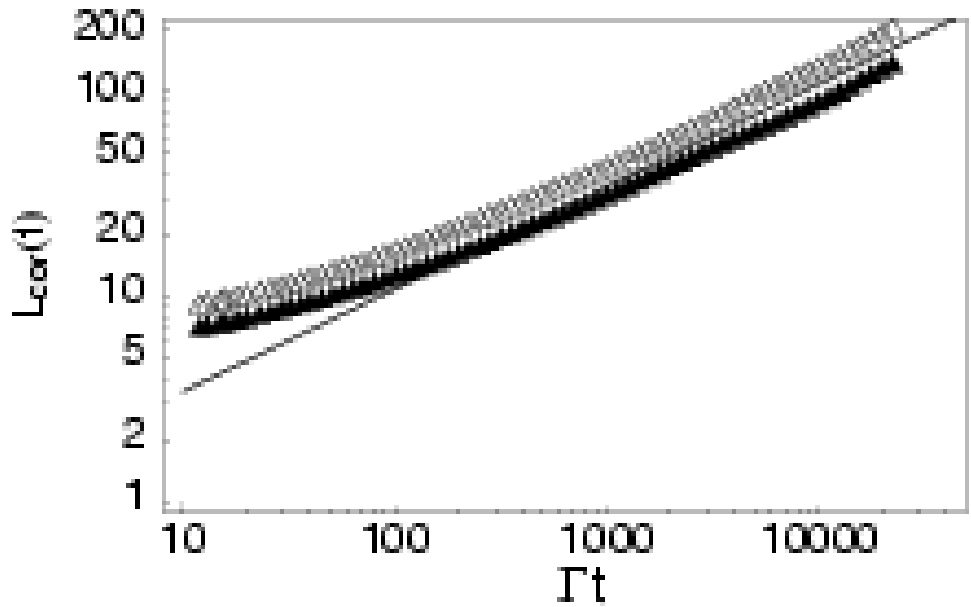}}
\vskip 0.2true cm
\caption{Length scale obtained from the correlation function as
  a function of time.  The eyeguide line has slope $1/2$. The symbols
  correspond to empty triangle: $\gamma=3.25$ and 
filled triangle: $\gamma=3.00$.  }
\label{Lcor512}
\end{figure}

\begin{figure}
\centerline{\epsfxsize=5.in
\epsffile{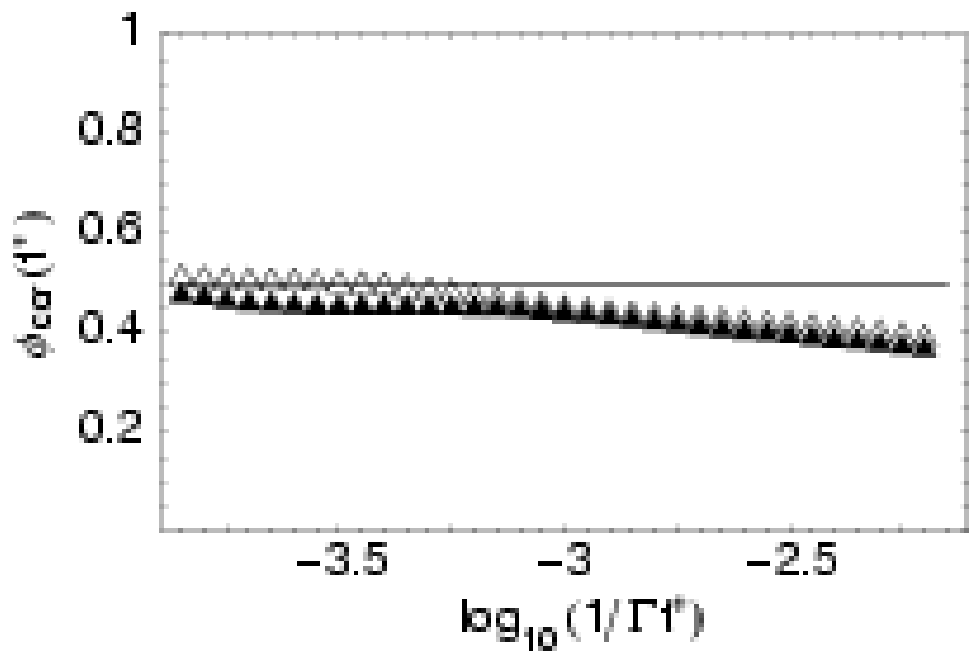}}
\vskip 0.2true cm
\caption{Effective exponent for the growth of the correlation length
  as a function of time: filled triangle: $\gamma=3.00$, empty triangle:
  $\gamma=3.25$.}
\label{phicor512}
\end{figure}

\begin{figure}
\centerline{\epsfxsize=5.in
\epsffile{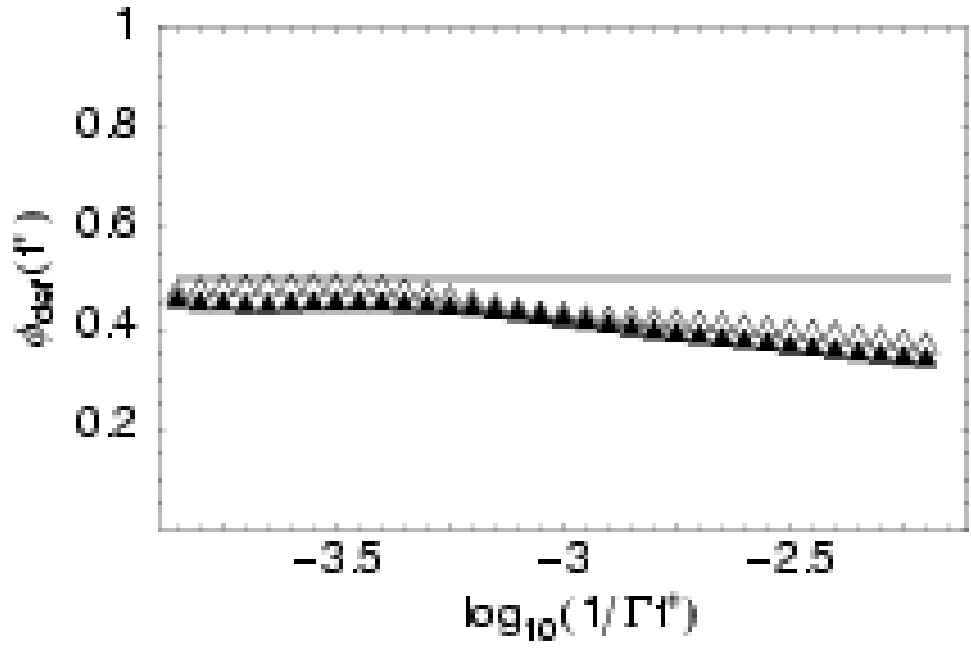}}
\vskip 0.2true cm
\caption{Effective exponent for the growth of the defect separation length
  as a function of time filled triangle: $\gamma=3.00$, empty triangle:
  $\gamma=3.25$.}
\label{phidef512}
\end{figure}
\newpage

\begin{figure}
\centerline{\epsfxsize=5.in
\epsffile{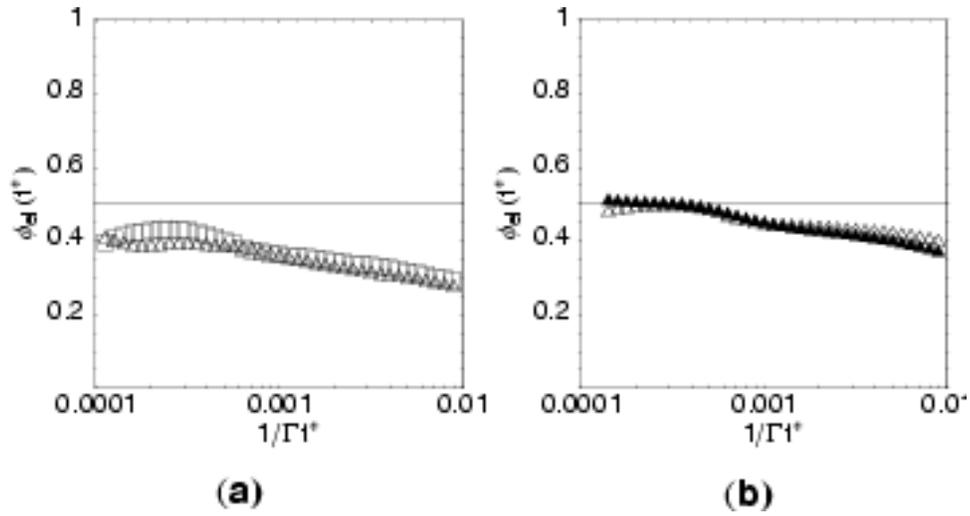}}
\vskip 0.2true cm
\caption{(a)Effective exponent for the growth of the elastic energy length
  as a function of time triangle: $\gamma=3.00$, box: $\gamma=3.25$.
  (b) Effective exponent for the correlation length (filled triangle),
  and the growth of the elastic energy
  corrected for the logarithmic scaling of Eq.(\ref{phid}) (empty triangle).}
\label{phien512}
\end{figure}
\newpage

\begin{figure}
\centerline{\epsfxsize=5.in
\epsffile{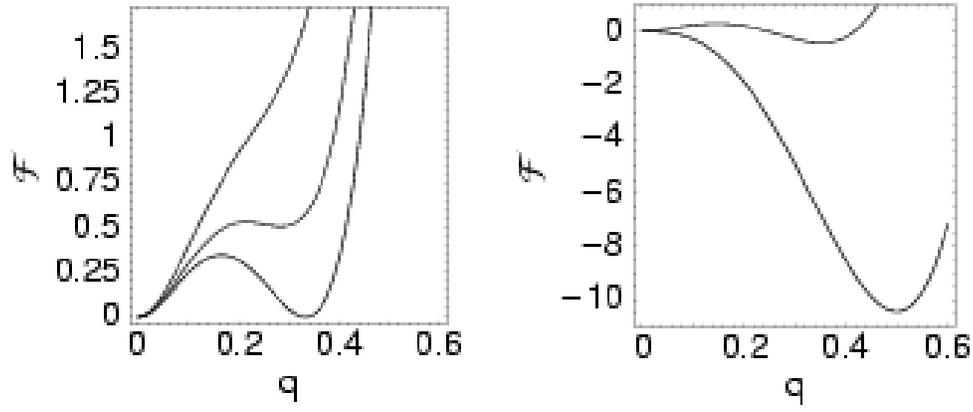}}
\vskip 0.2true cm
\caption{Free energy ($\times 10^3$) versus magnitude of the order
  parameter $q$ for $\gamma=2.7$ (left) and $\gamma=3.0$ (right).  Solid
  lines correspond to the assumption of complete orientational order
  and dashed lines correspond to the spatially averaged free energy for
  different values of the defect separation (6 and 12 in left figure and 2
  in right figure).} 
\label{freenwor}
\end{figure}

\begin{figure}
\centerline{\epsfxsize=5.in
\epsffile{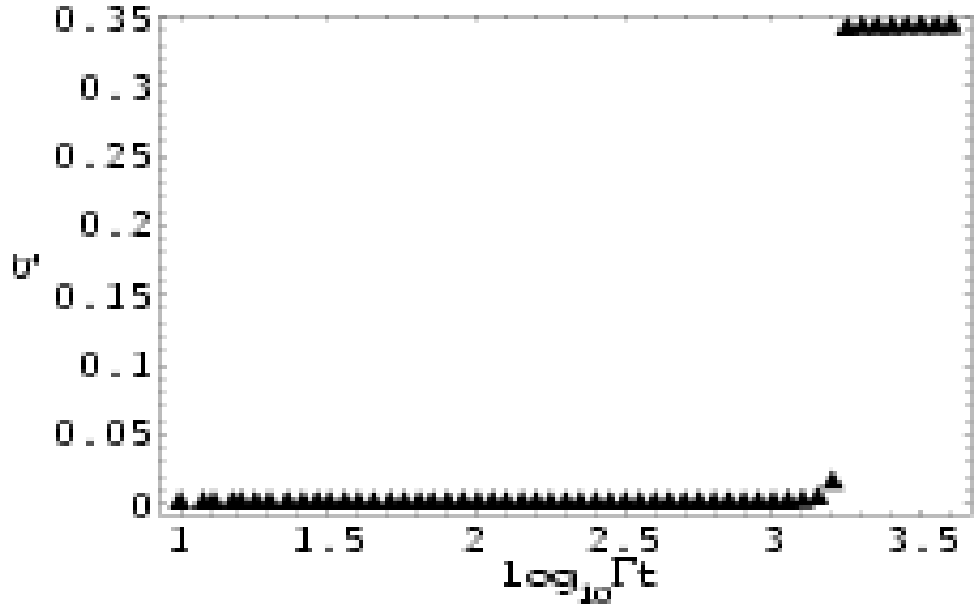}}
\vskip 0.2true cm
\caption{Amplitude $q$ versus time for a quench to $\gamma=3.05$. The
  initial fractional noise on $q$ ($\sim 10^{-6}$) is very small and
  cannot cause ordering by nucleation and growth.} 
\label{ampvst}
\end{figure}

\begin{figure}
\centerline{\epsfxsize=5.in
\epsffile{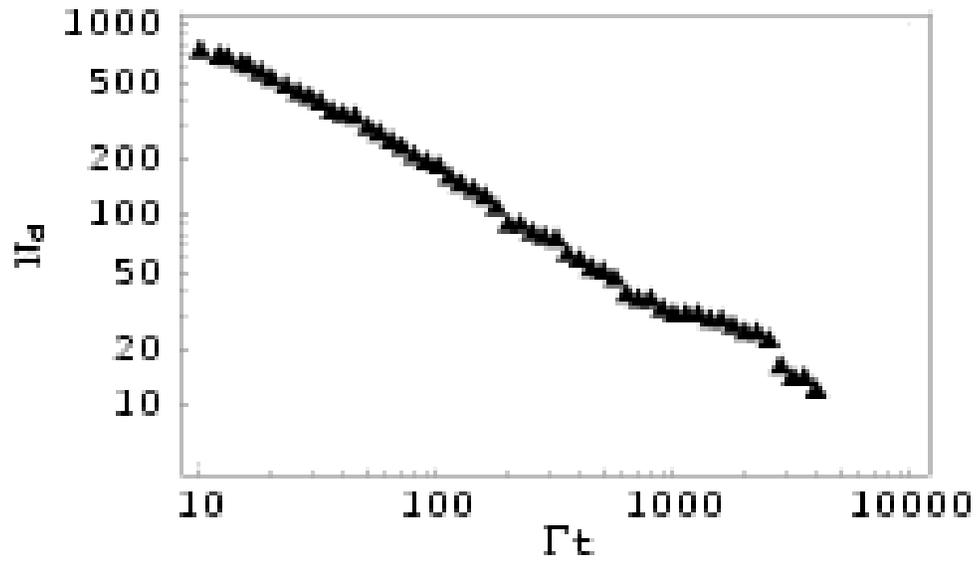}}
\vskip 0.2true cm
\caption{Number of defects versus time for a quench to $\gamma=3.05$.} 
\label{Ndvst}
\end{figure}

\end{document}